\documentclass{article}

\usepackage[english]{babel}

\usepackage[letterpaper,top=2cm,bottom=2cm,left=3cm,right=3cm,marginparwidth=1.75cm]{geometry}

\usepackage{amsmath}
\usepackage{graphicx}
\usepackage[colorlinks=true, allcolors=blue]{hyperref}
\usepackage{amsfonts}
\usepackage{booktabs}
\usepackage{siunitx}

\usepackage{caption}
\usepackage{subcaption}

\usepackage[table]{xcolor}

\usepackage[skip=0pt plus1pt, indent=15pt]{parskip}

\title{Assessing survival models by interval testing with Poisson-binomial distributions}
\author{Ben Lee}

\begin{document}

\maketitle

\begin{abstract}
Selecting appropriate parametric survival models is often a pivotal part of a regulatory submission for new pharmaceutical products. With recent developments in complex survival approaches, the number of suitable models is increasing, making model selection more challenging. Common approaches to model selection include AIC, BIC, and expert opinion on survival extrapolation. However, these approaches primarily assess relative goodness-of-fit, providing limited insight into where, and to what extent, a fitted model is incompatible with the observed data. 
We propose evaluating survival models using Poisson-binomial distributions across specified time intervals.  
Two interval selection approaches, censor-defined intervals and 10 evenly-spaced intervals, are presented with worked examples. A simulation exercise, targeting two proposed test statistics across 12 standard scenarios (with different data maturity and patient numbers), demonstrated that for every scenario the empirical Type I error did not exceed the nominal 5\% level. Our proposed model selection technique goes beyond classical approaches by highlighting time intervals where models perform poorly.
\end{abstract}

\


\

\noindent
An R-package for these methods, including examples is available \href{https://github.com/Ben-L-Stats/BITsurv}{here.}

\section{Introduction}\label{sec1}

Right-censored survival data are frequently collected in clinical trials and often form the basis of a trial's primary conclusions. To model the efficacy of the treatment within and beyond the duration of such trials, survival models can be fitted to the data. A choice between these fitted survival models must then be made, and this is often the driving factor in the economic evaluation of a new medical product.

\

\noindent
There are numerous approaches to fitting survival models.
Standard parametric survival models are often required for health technology assessment \cite{TSD14}.
There is also a growing emphasis on the use of flexible survival models, such as splines \cite{TSD21}.
In addition to this, complex approaches incorporating external data can also be used \cite{TSD21}. 
With the growing number of suitable approaches, model selection in health technology assessment becomes more challenging.

\

\noindent
Recommendations on selecting the most appropriate survival models include subjective assessments of curve fit and clinical validity, the use of external data, as well as objective assessments using AIC and BIC \cite{TSD14}.
The AIC and BIC are particularly useful as they are objective measures and can be easily computed.
However, the AIC and BIC only enable relative comparisons of one fitted model to another, meaning insight into whether all of the considered models are incompatible with observed data is not provided by such tests.
Furthermore, these tests do not indicate the specific locations where a fitted model is performing poorly.
Our proposed approach provides information on specific areas of poor fit and allows insight beyond relative comparisons.

\

\noindent
Our proposed approach uses groupings based on specified time-intervals.
The idea to group based on time intervals was proposed by Akritas (1988) \cite{Akritas1988} and subsequently consolidated and developed by several authors including Hollander Peña (1992) \cite{Holl1992} and Bagdonavicius Nikulin (2011) \cite{BN2011}.
The approach of Akritas is based on chi-squared tests, which require assumptions of asymptotic normality.
However, our proposed approach differs from this as it utilises Poisson-binomial distributions, which do not require any assumptions of asymptotic normality.

\ 

\noindent
The proposed Poisson-binomial approach accounts for right-censoring by splitting time into intervals defined from one censor to the next.
These subintervals are enclosed by each specified interval, for example one of ten evenly-spaced intervals in time.
Events observed in each specified interval are governed by a Poisson-binomial distribution defined using the subintervals and the fitted survival model.
This distribution can be used in a variety of ways to investigate the validity of the fitted survival model.

\

\noindent
The paper is structured as follows: methods, application, simulation, and discussion. 
The methods section describes: the proposed approach with intervals based entirely on censoring, an extension to this with specified intervals, and methods for generating an overall p-value. 
The application section includes two worked examples.
The first example evaluates a simple exponential model, providing an introduction to the developed methods.  
The second example is more practical for health technology assessment, with seven parametric models being compared using the developed methods.
A simulation elucidates the validity of the proposed approach, with empirical Type I error rates being presented for various simulated scenarios.
Finally, a discussion is given.

\section{Methods}\label{sec2}

\subsection{Interval testing}

We consider right-censored survival data, where for each subject we have (time, event) $\in (\mathbb{R}_{> 0}, \{ 1,0 \})$. We split this data into two types:

\begin{itemize}
  \item The observed data (event=1) with times
$t_{i,1}$ such that $0<t_{i,1} \le t_{i+1,1}$ and $i=1,\ldots, E.$

  \item The censored data (event=0) with times
$t_{i,0}$ such that $0<t_{i,0}\le t_{i+1,0}$ and $i=1,\ldots, C.$
\end{itemize}

\noindent
Subsequently, we define the set of unique censor times $t_{j,0}'$ for $j=1,\ldots, J$. That is, the values $t_{j,0}'$ are the unique values from the set of $t_{i,0}$ values, and thereby the $t_{j,0}'$ values satisfy $0<t_{j,0}'<t_{j+1,0}'$. We now define the intervals of sequential censor times such that for $j=1, 2,..., J$

\begin{equation*}
I_j= \begin{cases} 
      (0, \ t_{1,0}']  \ \   \   \   \  \  \  \    \   for \  \  \ j=1 \\
      (t_{j-1,0}', \ t_{j,0}'] \ \ \  \  else
   \end{cases}
\end{equation*}

\

\noindent
We are interested in the hypothesis that the event generating process is described by the model $M(\boldsymbol{\beta})$, with probability density function $f_M$ and fixed parameters $\boldsymbol{\beta}$.
We now make the assumption that the intervals themselves, $I_j$, are uninformative of the event generating process.
It follows that under the null, the distribution of observed events within each interval can be approximated by
\begin{equation} \label{eq:1}
events(I_j) \sim binomial(N.risk(I_j),  \ p_{I_j}). 
\end{equation}

\

\noindent
$events(I_j)$ are the number of events observed within the interval $I_j$.

\

\noindent
$N.risk(I_j)$ are the number of patients that enter the interval $I_j$.

\

\noindent
$p_{I_j}=$ ``the probability of a subject experiencing an event in interval $I_j$ given the subject makes it to interval $I_j$", which is given by
\begin{equation} \label{eq:pI}
\begin{split}
    p_{I_j} &=\mathbb{P}(  T \in I_j  \ | \ T>min(I_j) \ ) \\
            &=\mathbb{P}(  T \in I_j   ) \ /  \ \mathbb{P}(  T>min(I_j)  ) \\
            &=\int_{I_j} f_M(t) dt \Big/    \int_{min(I_j)}^{\infty} f_M(t) dt.
\end{split}
\end{equation}

\

\noindent
In summary, for each interval $I_j$ we have specified a model with known $N.risk(I_j)$, $p_{I_j}$, and observed $events(I_j)$. 
Thereby, each interval has a (midpoint) p-value for observing $events(I_j)$ in that interval.

\subsection{Extension to specified intervals} \label{extension}

The methodology above describes how events can be modelled within each interval. However, the intervals themselves are entirely determined by the censor times. It is often of interest to understand events in specified intervals (for example 10 evenly-spaced intervals up to the last censor). Below, we extend the methodology to allow for such an analysis.

\

\noindent
Define the intervals of interest for $k=1, \dots, K$ as
\begin{equation*}
V_k= \begin{cases} 
      (S_0, \ S_{1}]   \ \   \   \   \  \  \  \    \   for \  \  \ k=1 \\
      (S_{k-1}, \ S_{k}] \ \ \  \  \  \     else
   \end{cases}
\end{equation*}

\noindent
where $0\le S_0<S_1< \dots <S_{K}$.

\ 

\noindent
We wish to define a unique set of times $\tau_i$.
We have unique censor times $t_{j,0}'$ for $j=1,\ldots, J$ and unique $S_k$ values for $k=0, \dots, K$, taking the unique values of the union of these sets, we obtain the values $0<\tau_i<\tau_{i+1}$.  
From these, we construct the $I'$ intervals for $i=1, 2,..., L_{\tau}$, where $L_{\tau}$ denotes the total number of $\tau$ values.
\begin{equation*}
I'_i= \begin{cases} 
      (0, \tau_1]   \ \   \   \   \  \  \  \    \   for \  \  \ i=1 \\
      (\tau_{i-1}, \tau_{i}] \ \ \  \  \  \     else
   \end{cases}
\end{equation*}

\noindent
As in Equation \ref{eq:1}, we now have a known approximate distribution for each interval $I_i'$. The observations in said intervals are described by
\begin{equation*} 
events(I_i') \sim binomial(N.risk(I_i'), \  p_{I_i'}), 
\end{equation*}

\noindent
with $events(I_i')$, $N.risk(I_i')$ and $p_{I_i'}$ defined analogously to Equations \ref{eq:1} and \ref{eq:pI}. 

\

\noindent
We now consider the $I_i'$ intervals as subintervals that make up each $V_k$ interval. For each $V_k$, $k=1, \dots, K$, we define the collection of $L_k$ subintervals with $m=1,..., L_k$ as
\begin{equation*} 
\{I_{m,V_k}'\}= \{I_i' \ \  \  such \ that   \ \ I_i' \subseteq V_k \}.
\end{equation*}

\noindent
It follows that for each interval $V_k$ its subintervals follow 
\begin{equation*} 
\begin{split}
events(I_{1,V_k}') &\sim binomial(N.risk(I_{1,V_k}'), \ p_{I_{1,V_k}'}) \\
...                                                                \\
events(I_{L_k,V_k}') &\sim binomial(N.risk({I_{L_k,V_k}'}), \ p_{I_{L_k,V_k}'}). \\
\end{split}
\end{equation*}

\noindent
Thereby $\sum^{L_k}_{m=1} events({I_{m,V_k}'})$ is distributed as a sum of binomials distributions with known parameters, which is a special case of the Poisson-binomial distribution. 
Thus, we can obtain a (midpoint) p-value for each specified interval $V_k$, according to the number of observed events in that interval.

\subsection{Test statistics} \label{TS}

The primary recommendation of this paper is the use of interval plots, such as Figure \ref{fig:1}b. 
To support these plots, a single overall p-value can be a useful tool. 
Such a p-value can be obtained by condensing the plot's constituent p-values into a single overall p-value.
When deriving these test statistics, we assume that the p-value for each interval follows a continuous uniform[0,1] distribution under H0.

\

\noindent
We derive two test statistics for obtaining an overall p-value. 
From Birnbaum (1954) \cite{Birn}, we know that selecting a best test statistic for H0 is not trivial, and a more modern discussion is given in \cite{combineps}. 
Our first test statistic is continuous and allows for extreme p-values to have a high influence on the overall p-value. The second test statistic is discrete and aims to mitigate the influence of extreme p-values.

\subsubsection*{Transformed Fisher test (TFT) statistic}

To motivate the test, we consider a p-value for a given interval. 
A high p-value indicates more events have occurred in the interval than would generally be expected under H0. 
A low p-value indicates fewer events than expected. 
It follows that both high and low p-values indicate poor model fit within a given interval. 
As such, we will construct a  test statistic that can detect an uncommon amount of high and low p-values.

\

\noindent
For each interval, we have a p-value, $p_j$. Under H0, we make the assumption these p-values are $uniform[0,1]$. The p-values are transformed to the form
\[ U_j= \begin{cases} 
      2p_j \ \ \ \ \  \  \ \ \ \  \ for \ \  p_j \le 0.5 \\
      2(1-p_j)  \ \ \ \ for \ \  p_j > 0.5
   \end{cases}
\]

\noindent
It follows that $U_j \sim uniform[0,1]$. By the classical result from Fisher, we obtain the test statistic

\begin{equation*}
    T_{cont}=-2\sum_{j=1}^{I}log(U_j) \sim \chi^2_{_{2I}}, 
\end{equation*}

\noindent
where we reject for large $T_{cont}$ and $I$ is the total number of intervals. The intuition for this test is that unfavourable p-values (close to 0 or 1) are transformed such that they correspond to $U_j$ values close to 0. This makes the overall test able to detect unfavourable p-values.

\subsubsection*{Protection Against Very Small Intervals (PAVSI) test statistic}

The transformed Fisher test is useful as it enables us to leverage the full range of reported p-values. 
However, in some situations, this may not be ideal. 
For example, a single p-value that is very close to 0 can heavily influence the test. 
We argue that a test that is not dominated by extreme p-values is a useful tool.

\

\noindent
For each interval, we have a p-value, $p_j$. Under H0, we make the assumption these p-values are $uniform[0,1]$. The p-values are transformed to the form

\ 
\[ W_j= \begin{cases} 
      1 \ \ \ for \ \  p_j \ge 0.975  \ \  or \ \ p_j \le 0.025 \\
      0 \ \ \ else
   \end{cases}
\]

\noindent
It follows that under H0 we have $W_j \sim bernoulli(0.05)$, and thereby 

\begin{equation*}
    T_{pavsi}=\sum_{j=1}^{I}W_j\sim binom(I, 0.05), 
\end{equation*}

\noindent
where we reject for large $T_{pavsi}$ values. 
It is noted that $T_{pavsi}$ is a discrete test statistic.
Therefore, in Examples 1 and 2, the PAVSI is calculated using the midpoint p-value approach.

\ 

\noindent
If the reader is interested in using PAVSI in there own work, there are some technical points to consider that are discussed in Appendix \ref{PAVSI}.

\newpage

\section{Application}

The methods above described two general approaches to interval testing:
\begin{itemize} 
\item Use the censor times to specify the intervals
\item Specify the intervals yourself (``extension to specified intervals")
\end{itemize}

\noindent
In example 1, we focus on the first approach only. 
In example 2, both approaches are described.
For examples 1 and 2, we use real world overall survival data for melanoma from Freeman et al \cite{Freeman}. 
Example 1 uses the `Dabrafenib' treatment arm of the BREAK-3 trial \cite{BREAK3}. 
Example 2 uses the `Dabrafenib+Trametinib' treatment arm of the COMBI-d trial \cite{COMBI-d}.
All data and code is available on Github \href{https://github.com/Ben-L-Stats/BITsurv}{here.}

\subsection{Example 1: A fitted exponential model}

\begin{figure*}[ht]
\centerline{\includegraphics[width=0.65\linewidth]{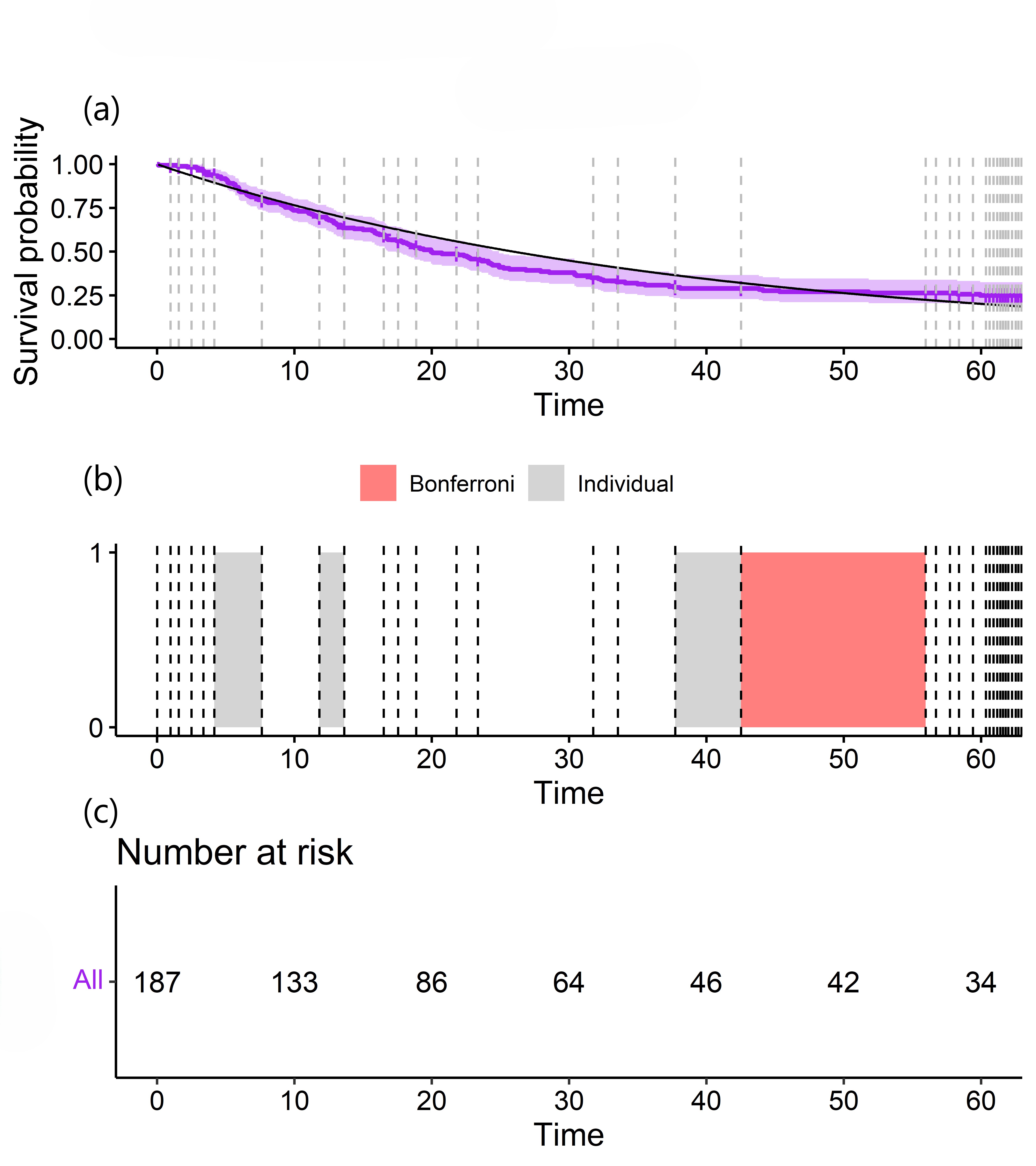}}
\caption{Example 1: Plot results. \textbf{Part a:} The KM curve for the data in purple, censors are given by dashed vertical black lines, and the fitted exponential curve is the overlaid curve in black. 
    \textbf{Part b:} The interval tests for this data. Intervals are defined by the censors, where for each interval, red indicates a Bonferroni rejection, and grey indicates an individual flag. 
    \textbf{Part c:} Numbers at risk for the data} \label{fig:1}
\end{figure*}

In this example, we fit the simplest survival model (exponential) to our data. 
Following this, we want to determine how appropriate the exponential model is.
First, we do a visual check of the curve fit, Figure \ref{fig:1}a. 
Visually, the Kaplan-Meier curve and the exponential curve seem to be misaligned, suggesting poor fit.
However, an interval test provides a more rigorous assessment of this, Figure \ref{fig:1}b.
For this example, we are only interested in an interval test where the intervals are selected using the censor times. 

\

\noindent
To perform the test, we first obtain the values $I_j$, $N.risk(I_j)$, and $events(I_j)$ directly from the data. 
To complete equation \ref{eq:1}, we must derive $p_{I_j}$ for the fitted exponential model.
As the probability density function for the exponential distribution is given by $f_M(t)=\lambda e^{-\lambda t}$, using Equation \ref{eq:pI}, it follows that
\begin{equation*}
p_{I_j} =1- exp(\lambda [min(I_j)-max(I_j) ]  \ ).
\end{equation*}

\noindent
The value $\lambda$ is known (from fitting the exponential model to the data). We have therefore fully specified the binomial realisations for each interval $I_j$. This information for the first five intervals is presented in Table \ref{tab:111}.

\

\noindent
The p-value for each interval is discrete as it comes from a binomial model. It follows that each discrete p-value does not follow a continuous uniform[0,1] distribution under H0. This can be rectified by using randomised p-values \cite{midps}. However, in the interest of reproducibility and clarity, we will focus our discussion on the results of the midpoint p-values. (This is discussed in more detail in Appendix \ref{midp}.)

\ 

\noindent
For each interval, we consider the midpoint p-value with respect to a 2-sided test. 
That is, if $p_{I_j}\le 0.025$ or $p_{I_j}\ge 0.975$, we have reason to suspect the interval generating said value. 
We call this the individual test.
Note that this is an informal tool, and we do not accept or reject a model on the basis of a single individual test.

\ 

\noindent
Of greater concern, are the intervals that fail the Bonferroni test \cite{Bonferroni}. 
The Bonferroni test is chosen to strongly control the familywise error rate at a level of 0.05. 
In particular, we reject if $p_{I_j}\le 0.025/I$ or $p_{I_j}\ge 0.975/I$, where $I=$``the total number of intervals". 
This means that under H0 (that the model holds for all intervals) a single Bonferroni rejection of an interval is only expected to occur for 1 in 20 times the data is generated under H0.

\ 

\noindent
In Figure \ref{fig:1}, we present the results for the Bonferroni and individual tests alongside the Kaplan-Meier (KM) curve, the fitted exponential model, and the numbers at risk.
A single Bonferroni rejection of an interval is good reasoning for rejecting a model. 
In Figure \ref{fig:1}, the very poor fit (Bonferroni rejection) is at the end of the curve. 
This is very unfavourable in health economic evaluation, where plausible extrapolations beyond the observed data are paramount to model selection. 
That is, the Bonferroni rejection at the end of the curve indicates that the extrapolation beyond the trial period may be poor.

\newpage

\begin{table*}[!t]%
\centering %
\caption{Example 1: Summary of interval data. \label{tab:111}}%

\begin{tabular*}{\textwidth}{@{\extracolsep\fill}llllll@{\extracolsep\fill}}
\toprule
{$I_j$} & {N.risk} & {$p_{I_j}$} & {$events(I_j)$} & {$\mathbb{E}$(events)} & {p-mid} \\

\midrule
 (0, 1]     & 187 & 0.026 & 1 & 4.8 & 0.027  \\ 
 (1, 1.6]   & 185 & 0.016 & 1 & 3.0 & 0.127  \\ 
 (1.6, 2.5] & 183 & 0.024 & 1 & 4.4 & 0.037 \\ 
 (2.5, 3.4] & 181 & 0.023 & 4 & 4.2 & 0.491  \\ 
 (3.4, 4.2] & 176 & 0.021 & 5 & 3.6 & 0.771  \\ 
 ... & ... & ... & ... & ... & ... \\

\bottomrule
\end{tabular*}
\end{table*}

\ 

\noindent
An overall p-value for the fitted exponential model is also of interest. 
For the transformed Fisher test statistic, there are $I=43$ intervals, and our test statistic was computed as $-2\sum_{j=1}^{I}log(U_j)=81.84$, giving a p-value of 0.607.
For the PAVSI test statistic, there are $I=43$ intervals. 
From Figure \ref{fig:1}, we can read off that $T_{pavsi}=4$ (as there are 3 individual rejections and 1 Bonferroni rejection).  
This gives us an overall midpoint p-value of 0.114.
(These test statistics should be used with caution as our simulation indicated that they are somewhat conservative in practice.)

\ 

\noindent
As we can see, the two test statistics yield very different results. 
It is our recommendation that a model should not be accepted or rejected based entirely on the outcome of such tests. 
Instead, they are tools that provide an initial overview of model fit. 
The greatest detail and insight can be obtained by assessing the binomial interval plot alongside specific interval p-values and an overall p-value.
For this particular example, we would recommend that the model is rejected.
This is primarily due to the Bonferroni rejection at the end of the curve.

\subsection{Example 2: Selecting a parametric survival model for challenging data}

In this example, the survival data has a challenging shape, making it difficult for models to provide good visual fit.
We fit seven standard parametric survival models using flexsurv \cite{flexsurv}.
Our results and tables are then generated by the BITsurv package, which is available \href{https://github.com/Ben-L-Stats/BITsurv}{here.}
Our aim in Example 2 is to determine whether any of the seven survival models are appropriate, or whether survival models with additional flexibility are required.
Both censor interval and specified interval approaches are used.

\subsubsection{Intervals selected based on censors}

First, we perform the interval test with intervals chosen using censors as in Example 1.
The 7 parametric models were fitted, and the test statistics for each parametric model are presented in Table \ref{tab2}.
The PAVSI and TFT overall p-values are well above the 0.05 threshold for all of the survival models.
However, as will be discussed in our simulation section, these test statistics are often conservative for the censor interval approach.

\

\noindent
For each model, there are between 1 and 4 individual flags.
However, these flags are primarily descriptive.
Of greater interest are the Bonferroni rejections, of which there are none.
This seems good, and from Table \ref{tab2} alone we might believe that our curves provide good fit.
However, this is an unfortunate result of using intervals based on censors.
Due to the large number of censors, the majority of intervals are very small, which makes it difficult to detect poor fit within these intervals.
For example, there are many small intervals for the last 5 months of the curve.
Instead, it would be more informative if we could consider a single interval for those final 5 months.
To allow this, the specified interval approach is used.

\begin{table}[h!] 
\begin{center} 
\caption{Example 2: Interval test results using censor defined intervals.} \label{tab2}
\begingroup
\renewcommand*{\arraystretch}{1.3}

\begin{tabular}{||c c c c c||} 
 \hline
 {Parametric survival model}  & {Bonferroni rejections} & {Individual flags} & {PAVSI} & {TFT}  \\ [0.5ex] 
 \hline\hline
Exponential       & 0 & 3 & 0.468 & 0.854  \\
 \hline
Gamma             & 0 & 2 & 0.696 & 0.887 \\ 
 \hline
Generalised gamma & 0 & 1 & 0.881 & 0.955 \\ 
 \hline
Gompertz          & 0 & 4 & 0.267 & 0.884 \\ 
 \hline
Log-logistic      & 0 & 1 & 0.881 & 0.977 \\ 
 \hline
Log-normal        & 0 & 2 & 0.696 & 0.968 \\ 
 \hline
Weibull           & 0 & 3 & 0.468 & 0.871 \\ 
 \hline
\end{tabular}
\endgroup
\end{center}
\end{table}

\subsubsection{Ten evenly-spaced intervals (a specified interval approach)}

There are many ways these intervals could be chosen, and several potential approaches are briefly mentioned in the discussion. 
Here, we will outline one of the simpler approaches.
The data is split into 10 evenly-spaced intervals. 
These intervals start at time 0 and end at the largest censor time. That is, for $k=1,...,10$

\begin{equation*}
    V_k=\frac{1}{10} \cdot t.max \cdot (k-1, \ k].
\end{equation*}

\begin{figure*}
\centerline{\includegraphics[width=0.7 \linewidth]{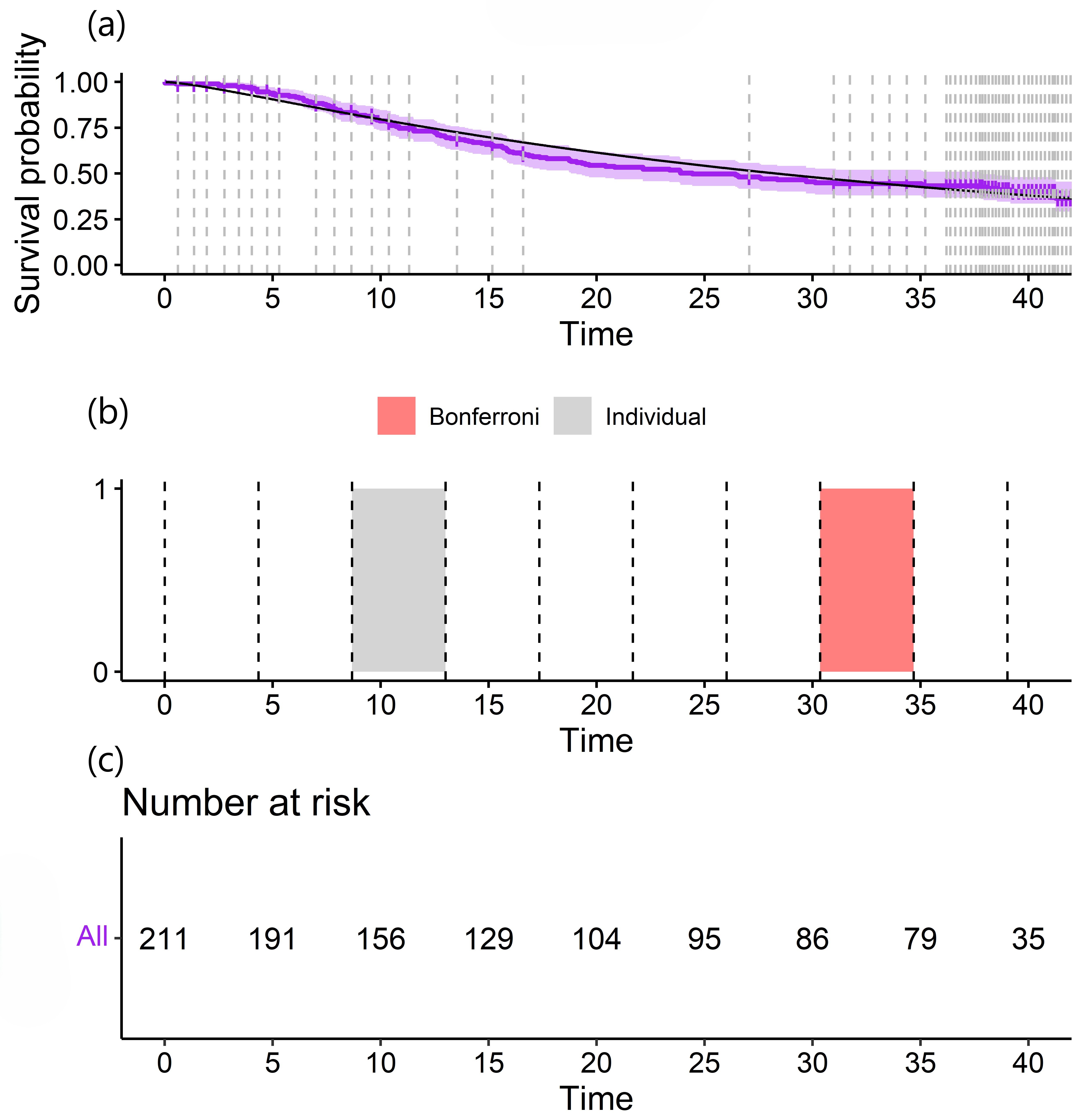}}
\caption{Example 2: Plot results for generalised gamma. \textbf{Part a:} The KM curve (purple), censors (vertical dashed lines), and the fitted generalised gamma curve (black). 
             \textbf{Part b:} The interval test results for 10 evenly-spaced intervals. 
             \textbf{Part c:} Numbers at risk.}
             \label{gengamma}
\end{figure*}

\noindent
Applying this approach to the data yields the results presented in Table \ref{tab3}.
Of the seven fitted curves, we present the generalised gamma results in Figure \ref{gengamma}.

\

\noindent
The results of Table \ref{tab3} tell a very different story to the censor interval approach.
First, the overall TFT p-value is below 0.05 for all models except log-normal, indicating poor fit in 6 out of 7 models.
Similarly, the overall PAVSI p-value mostly agrees with the TFT, rejecting 5 out of 7 models.
However, the PAVSI should be interpreted with caution when using the 10 evenly-spaced interval approach as it can inflate the Type I error rate, see discussion Appendix \ref{PAVSI}.

\

\noindent
Finally, all of the parametric models have a Bonferroni rejection except log-normal.
Because of this, we investigate the log-normal model in further detail, reviewing the results of each of its 10 intervals.
From Table \ref{tab4}, we see that the only interval resulting in an individual flag for the log-normal model was between times 30.35 and 34.69 with a p-value of 0.0057, which is above the 0.0025 threshold of a Bonferroni rejection.
Furthermore, we note that the 4th interval had a p-value of 0.9710. 
If this value was only 0.004 higher, then the log-normal would have received a second individual flag, and as a result of this the PAVSI overall p-value would have yielded a rejection (noting caveat Appendix \ref{PAVSI}).
For these reasons, as well as the overall TFT p-value of 0.0586 (relatively close to 0.05) we can be justifiably suspicious of the log-normal model.
However, we have no strong evidence to outright reject it.

\

\noindent
In conclusion, due to the high number of Bonferroni and TFT rejections, we suggest that the majority of the presented models are unsuitable.
In terms of the interval tests, the log-logistic and log-normal were the best performers.
However, even these models had their limitations, with the log-normal model being the worst of all 7 models for both AIC and BIC (not presented).
For these reasons, it is recommended that exploring flexible models such as splines and piecewise would be appropriate for this data.

\begin{table}[h!] 
\begin{center} 
\caption{Example 2: Interval test results using 10 evenly-spaced intervals.} \label{tab3}
\begingroup
\renewcommand*{\arraystretch}{1.3}
\begin{tabular}{||c c c c c||} 
 \hline
 {Parametric survival model}  & {Bonferroni rejections} & {Individual flags} & {PAVSI} & {TFT}  \\ [0.5ex] 
 \hline\hline
Exponential&       \cellcolor[HTML]{fc8672}1 & 4 & \cellcolor[HTML]{fc8672}0.0005 &\cellcolor[HTML]{fc8672}0.0029 \\ 
 \hline
Gamma&             \cellcolor[HTML]{fc8672}1 & 2 & \cellcolor[HTML]{fc8672}0.0488 &\cellcolor[HTML]{fc8672}0.0036 \\ 
 \hline
Generalised gamma& \cellcolor[HTML]{fc8672}1 & 2 & \cellcolor[HTML]{fc8672}0.0488 &\cellcolor[HTML]{fc8672}0.0261 \\ 
 \hline
Gompertz&          \cellcolor[HTML]{fc8672}1 & 4 & \cellcolor[HTML]{fc8672}0.0005 &\cellcolor[HTML]{fc8672}0.0042 \\ 
 \hline
Log-logistic&      \cellcolor[HTML]{fc8672}1 & 1 &                         0.2437 &\cellcolor[HTML]{fc8672}0.0491 \\ 
 \hline
Log-normal&                                0 & 1 &                         0.2437 &\cellcolor[HTML]{f3aa7f}0.0586 \\ 
 \hline
Weibull&           \cellcolor[HTML]{fc8672}1 & 3 & \cellcolor[HTML]{fc8672}0.0063 &\cellcolor[HTML]{fc8672}0.0029 \\ 
 \hline
\end{tabular}
\endgroup
\end{center}
\end{table}

\begin{table*}[!h]%
\centering %
\caption{Example 2: Interval test details for the log-normal model with 10 evenly-spaced intervals.\label{tab4}}%

\begin{tabular*}{\textwidth}{@{\extracolsep\fill}llllll@{\extracolsep\fill}}
\toprule
{$V_j$}  & {$\mathbb{E}$(Events)} & {Obs Events} & {p-mid} & {Individual test} & {Bonferroni test} \\

\midrule
(0, 4.34]     & 18.8 & 11& 0.0289 &Accept&Accept \\ 
(4.34, 8.67]  & 22.7 & 23& 0.5411 &Accept&Accept\\ 
(8.67, 13.01] & 17.9 & 26& 0.9677 &Accept&Accept \\ 
(13.01, 17.34]& 13.7 & 21& 0.9710 &Accept&Accept \\ 
(17.34, 21.68]& 10.7 & 11& 0.5549 &Accept&Accept \\ 
(21.68, 26.01]& 9.0  & 7 & 0.2550 &Accept&Accept \\ 
(26.01, 30.35]& 7.7  & 9 & 0.6947 &Accept&Accept \\ 
(30.35, 34.69]& 6.6  & 1 & 0.0057 &Flag&Accept \\ 
(34.69, 39.02]& 5.1  & 5 & 0.5156 &Accept&Accept \\ 
(39.02, 43.36]& 1.7  & 3 & 0.8280 &Accept&Accept \\ 

\bottomrule
\end{tabular*}
\end{table*}

\section{Simulation}

We will now pay particular attention to the properties of the proposed approach under different simulated scenarios. Specifically, we focus on the Type I error, where we aim to show that this is appropriately controlled at the nominal rate.

\subsection{Construction of simulation}

We focus our attention on the exponential model as this allows us scope to explore other critical scenarios such as data maturity and patient numbers. Applications to additional scenarios (such as other parametric models) could be explored in future work.

\

\noindent
Each simulation is described by the underlying event generation process: $T \sim exp(\lambda)$ and an underlying censor generating process for $C$ given by
\begin{equation*} 
\begin{split}
&C1 \sim uniform[0,100], \\
&C2 \sim uniform[18,22], \\
&C = min(C1, C2).
\end{split}
\end{equation*}

\noindent
The observable trial data is constructed as $Time=min(C,T)$; and $Event=1$ if $T\le C$, else $Event=0$. This construction can be visualised as a 22 Month trial with heavy censoring in the last 4 months of the trial and infrequent censoring throughout.

\subsection{Scenarios and implementation}

\noindent
In total, 12 data scenarios will be explored. For the number of patients in the fitted treatment arm, we consider 4 possible sizes, $N: 50, 100, 200$ and $500$. For data maturity, we consider three levels: $\lambda=1/10$ (mature), $\lambda=1/30$ (moderate), and $\lambda=1/70$ (immature), where at the end of the trial period (22 months), these levels of maturity imply underlying survival rates of $0.11, 0.48,$ and $0.73$ respectively. 

\

\noindent
For all 12 scenarios, 10,000 simulations were run for each interval method (10 evenly-spaced and censor) and each p-value approach (randomised and midpoint), totaling 480,000 simulated trial datasets.

\

\noindent
The empirical Type I error rates were calculated for each scenario as a proportion of the number of rejections among the 10,000 iterations.
The rejections were evaluated at the nominal 0.05 level.
Specifically, for the TFT, a rejection occured if the TFT test statistic was < 0.05.
For Bonferroni, a rejection occured if 1 or more intervals received a Bonferroni rejection, where the Bonferroni test was constructed to strongly control the familywise error rate at the 0.05 level. 
PAVSI was not presented; this was primarily to simplify the results, but also due to some minor limitations described in Appendix \ref{PAVSI}.

\

\noindent
Regarding the uncertainty associated with the simulation results, our empirical Type I errors can be represented by the approximate 95\% Wald confidence intervals for binomial-based proportions. 
We do not present these confidence intervals in the results tables to avoid overcrowding.
However, an empirical Type I error of 0.05 yields an associated 95\% confidence intervals of (0.0457, 0.0543). 
This confidence interval is relatively slim, due to the large number of simulations (10,000), indicating that our simulation results are generally reliable for proportions close to 0.05.
With the exception of Table \ref{tabA1}, the majority of our results yield proportions between 0.02 and 0.05, with 0.02 yielding a Wald 95\% confidence interval of (0.0173, 0.0227).

\subsection{Simulation results: randomised p-values}

\noindent
Randomised p-values allow us to test our underlying model assumptions via a simulation.
In particular, if the model assumptions are accurate, randomised p-values are well behaved as they should follow uniform[0, 1] distributions under H0.
On the other hand, midpoint p-values do not follow such a predictable distribution, and thereby we handle this case in a separate section.

\

\noindent
For all scenarios that used randomised p-values, both the 10 evenly-spaced interval and censor interval approaches behaved well with respect to Type I error rates. 
That is, the empirical Type I error rates were close to the nominal 0.05 level (between 0.040 and 0.055 in all cases) for both the Bonferroni and TFT tests, see Tables \ref{tabD1} and \ref{tabD2}. 
This indicates that the model assumptions holds well in all of our simulated scenarios for both the 10 evenly-spaced interval and censor interval approaches.

\begin{center}
\begin{table*}[!h]%
\caption{Empirical Type I error rates - Simulation results: Censor interval approach, randomised p-values\label{tabD1}}
\begin{tabular*}{\textwidth}{@{\extracolsep\fill}lllllll@{}}
\toprule
&\multicolumn{2}{@{}l}{\textbf{Mature}   }
& \multicolumn{2}{@{}l}{\textbf{Moderate} }
& \multicolumn{2}{@{}l}{\textbf{Immature}  } \\
&\multicolumn{2}{@{}l}{$\boldsymbol{\lambda =1/10}$    }
& \multicolumn{2}{@{}l}{$\boldsymbol{\lambda =1/30}$  }
& \multicolumn{2}{@{}l}{$\boldsymbol{\lambda =1/70}$  } 
\\\cmidrule{2-3}\cmidrule{4-5}\cmidrule{6-7}
\textbf{Number of patients} & 
\textbf{Bonferroni}  & \textbf{TFT}  &
\textbf{Bonferroni}  & \textbf{TFT}  &
\textbf{Bonferroni}  & \textbf{TFT}  \\
\midrule
50  & 0.0405 &	0.0428  & 0.0479 & 0.0524 & 0.0519 & 0.0549   \\
100 & 0.0431 &  0.0454  & 0.0524 & 0.0526 & 0.0438 & 0.0500   \\
200 & 0.0459 &	0.0467  & 0.0472 & 0.0499 & 0.0501 & 0.0494   \\
500 & 0.0493 &	0.0480  & 0.0486 & 0.0485 & 0.0488 & 0.0509   \\
\midrule
\multicolumn{7}{@{}l}{For each of the 12 scenarios (data maturity and patient numbers), 10,000 datasets were simulated.} \\
\bottomrule
\end{tabular*}
\end{table*}
\end{center}

\begin{center}
\begin{table*}[!h]%
\caption{Empirical Type I error rates - Simulation results: 10 evenly-spaced interval approach, randomised p-values\label{tabD2}}
\begin{tabular*}{\textwidth}{@{\extracolsep\fill}lllllll@{}}
\toprule
&\multicolumn{2}{@{}l}{\textbf{Mature}   }
& \multicolumn{2}{@{}l}{\textbf{Moderate} }
& \multicolumn{2}{@{}l}{\textbf{Immature}  } \\
&\multicolumn{2}{@{}l}{$\boldsymbol{\lambda =1/10}$    }
& \multicolumn{2}{@{}l}{$\boldsymbol{\lambda =1/30}$  }
& \multicolumn{2}{@{}l}{$\boldsymbol{\lambda =1/70}$  }
\\\cmidrule{2-3}\cmidrule{4-5}\cmidrule{6-7}
\textbf{Number of patients} & 
\textbf{Bonferroni}  & \textbf{TFT}  &
\textbf{Bonferroni}  & \textbf{TFT}  &
\textbf{Bonferroni}  & \textbf{TFT}  \\
\midrule
50  & 0.0481 & 0.0454  & 0.0505 & 0.0524 & 0.0502 & 0.0525   \\
100 & 0.0461 & 0.0452  & 0.0496 & 0.0480 & 0.0506 & 0.0533   \\
200 & 0.0448 & 0.0494  & 0.0506	& 0.0539 & 0.0455 &	0.0530   \\
500 & 0.0475 & 0.0438  & 0.0470	& 0.0478 & 0.0470 & 0.0542   \\
\midrule
\multicolumn{7}{@{}l}{For each of the 12 scenarios (data maturity and patient numbers), 10,000 datasets were simulated.} \\
\bottomrule
\end{tabular*}
\end{table*}
\end{center}

\subsection{Simulation results: midpoint p-values}

\noindent
Unlike randomised p-values, the midpoint p-values are not uniform[0, 1] under H0. Instead, this is an assumption we must make for practical purposes, see discussion in Appendix A. 
Below, we investigate the impact this assumption has on empirical Type I error rates for our simulation.

\

\noindent
When midpoint p-values are used, both the 10 evenly-spaced interval and censor interval approaches were inclined to be conservative for all scenarios of our simulation. 
The 10 evenly-spaced interval approach was the better performing approach, with empirical Type I error rates being closer to the targeted rate of 0.05 for both the TFT and Bonferroni tests.
On the other hand, the censor interval approach had empirical Type I error rates that were much more conservative, in particular for the TFT the rates were often close to 0.

\

\noindent 
Table \ref{tabA1} presents the results for the censor interval approach.
The empirical Type I error rates for the Bonferroni tests were conservative for the censor interval approach, being between 0.01 and 0.02 in most cases. 
However,  the TFT was very conservative, having an empirical Type I error of 0.0001 in all the cases where $\lambda=1/30$ or $\lambda=1/70$. 
This likely occurs because a large proportion of each simulated dataset’s intervals had an expected number of events close to 0, and the majority of the observations in these intervals were 0. 
This results in an artificially high number of p-values close to 0.5 when using the midpoint p-value approach, influencing the TFT and causing the conservative results. 
(The 10 evenly-spaced interval approach does not suffer from this limitation.)

\begin{center}
\begin{table*}[!h]%
\caption{Empirical Type I error rates - Simulation results: Censor interval approach, midpoint p-values\label{tabA1}}
\begin{tabular*}{\textwidth}{@{\extracolsep\fill}lllllll@{}}
\toprule
&\multicolumn{2}{@{}l}{\textbf{Mature}   }
& \multicolumn{2}{@{}l}{\textbf{Moderate} }
& \multicolumn{2}{@{}l}{\textbf{Immature}  } \\
&\multicolumn{2}{@{}l}{$\boldsymbol{\lambda =1/10}$    }
& \multicolumn{2}{@{}l}{$\boldsymbol{\lambda =1/30}$  }
& \multicolumn{2}{@{}l}{$\boldsymbol{\lambda =1/70}$  }
\\\cmidrule{2-3}\cmidrule{4-5}\cmidrule{6-7}
\textbf{Number of patients} & 
\textbf{Bonferroni}  & \textbf{TFT}  &
\textbf{Bonferroni}  & \textbf{TFT}  &
\textbf{Bonferroni}  & \textbf{TFT}  \\
\midrule
50  & 0.0187 &	0.0066  & 0.0137 & 0.0001 & 0.0106 &	0.0001   \\
100 & 0.0231 &  0.0034  & 0.0151 & 0.0001 & 0.0109 &	0.0001   \\
200 & 0.0221 &	0.0002  & 0.0139 & 0.0001 & 0.0118 &	0.0001   \\
500 & 0.0213 &	0.0001  & 0.0122 & 0.0001 & 0.0101 &	0.0001  \\
\midrule
\multicolumn{7}{@{}l}{For each of the 12 scenarios (data maturity and patient numbers), 10,000 datasets were simulated.} \\
\bottomrule
\end{tabular*}
\end{table*}
\end{center}

\noindent
Table \ref{tabA2} presents the results for the 10 evenly-spaced interval approach.
The 10 evenly-spaced interval approach performed preferably to the censor approach, with results generally being less conservative and closer to the nominal 0.05 rate.
Across the scenarios, Bonferroni results had an empirical Type I error rate that was between 0.02 and 0.05.
The TFT empirical Type I error rates were generally between 0.01 and 0.05.
Across the scenarios, data maturity had a moderate impact on the results, with less mature data generally resulting in more conservative tests (lower empirical Type I error rate). 
Patient numbers had a more distinct impact, with lower patient numbers generally resulting in more conservative tests.
As patient numbers increased, the empirical Type I error rates appeared to approach the targeted level of 0.05.

\begin{center}
\begin{table*}[!h]%
\caption{Empirical Type I error rates - Simulation results: 10 evenly-spaced interval approach, midpoint p-values\label{tabA2}}
\begin{tabular*}{\textwidth}{@{\extracolsep\fill}lllllll@{}}
\toprule
&\multicolumn{2}{@{}l}{\textbf{Mature}   }
& \multicolumn{2}{@{}l}{\textbf{Moderate} }
& \multicolumn{2}{@{}l}{\textbf{Immature}  } \\
&\multicolumn{2}{@{}l}{$\boldsymbol{\lambda =1/10}$   }
& \multicolumn{2}{@{}l}{$\boldsymbol{\lambda =1/30}$  }
& \multicolumn{2}{@{}l}{$\boldsymbol{\lambda =1/70}$ }
\\\cmidrule{2-3}\cmidrule{4-5}\cmidrule{6-7}
\textbf{Number of patients} & 
\textbf{Bonferroni}  & \textbf{TFT}  &
\textbf{Bonferroni}  & \textbf{TFT}  &
\textbf{Bonferroni}  & \textbf{TFT}  \\
\midrule
50  & 0.0273 & 0.0209  & 0.0203 & 0.0155 & 0.0232 & 0.0059   \\
100 & 0.0353 & 0.0301  & 0.0313 & 0.0329 & 0.0206 & 0.0186   \\
200 & 0.0402 & 0.0364  & 0.0405	& 0.0383 & 0.0312 &	0.0322   \\
500 & 0.0476 & 0.0465  & 0.0436	& 0.0482 & 0.0398 & 0.0428   \\
\midrule
\multicolumn{7}{@{}l}{For each of the 12 scenarios (data maturity and patient numbers), 10,000 datasets were simulated.} \\
\bottomrule
\end{tabular*}
\end{table*}
\end{center}

\section{Discussion}

There are many approaches for assessing the suitability of survival models, with AIC and BIC being the most common objective assessments.
However, AIC and BIC only enable relative comparisons between survival models and do not indicate the specific locations of poor fit.

\

\noindent
We performed a simulation for our proposed approach, which showed that the Type I error rate was appropriately controlled, being either conservative or close to the targeted 0.05 level.
Specifically, the randomised p-value approach showed empirical Type I error rates close to 0.05 in all scenarios.
Although randomised p-values are useful for confirming the properties of our proposed approach, they are not recommended for use in practice due to their challenging interpretation.
Instead, midpoint p-values are better suited for real-world use.

\ 

\noindent
In the simulation, we compared two approaches that utilise the proposed method: censor defined intervals and 10 evenly-spaced intervals.
As we recommend midpoint p-values in practice, we focus on this case when comparing the two approaches.
The censor interval approach was highly conservative for midpoint p-values, being between 0.01 and 0.02 for Bonferroni and frequently close to 0 for the TFT.
The 10 evenly-spaced interval approach was less conservative, with empirical Type I error rates generally ranging from 0.02 to 0.05, where higher patient numbers yielded rates close to 0.05.
For this reason, the 10 evenly-spaced interval approach is generally recommended over the censor interval approach.
Furthermore, Example 2 provides a practical case study of why the 10-evenly spaced interval approach might be preferable in practice.

\

\noindent
Regarding further developments, although we focused on 10 evenly-spaced intervals and censor-based intervals, the proposed methodology is flexible to accommodate many interval selection approaches.
One promising approach would be interval selection based on all intervals having an equal expected number of events.
Another approach would be to collapse small intervals into larger ones when they do not contain sufficient information. 
Guidance on various approaches, especially in the presence of limited patient numbers, would be a beneficial area of research.

\

\noindent
The proposed methodology naturally extends to risk score models, which could be another avenue for further research. 
Such models allow for unequal risk scores for each patient, whereas our examples inherently assumed equal risk scores, which is a common feature of models used in health technology assessment.
The risk score approach would require some modifications to the proposed methodology, specifically using a general form of the Poisson-binomial distribution rather than the special case of this in Equation \ref{eq:1}.

\

\noindent
Finally, future work could also compare the proposed method with other calibration approaches via an extensive simulation.
The objective of such a simulation would be to compare the overall test statistics of our approach to other approaches with respect to power and Type I error rates across a range of scenarios.
The key competing method in this potential simulation would be the Akritas chi-squared test \cite{Akritas1988}, as it is also a time-interval based approach.
The simulation could also be extended to a wider class of calibration approaches, such as Hosmer Lemeshow type tests \cite{Demler}, if utilising the risk score model extension.
The key distinction between Akritas and our own approach is that Akritas relies on assumptions of asymptotic normality, whereas our approach utilises Poisson-binomial distributions but relies on the assumption that the midpoint p-values behave as uniform[0,1] random variables.
For both approaches, increasing the population size improves the validity of these assumptions.

\

\noindent
In conclusion, we have presented an approach to assist in the assessment of survival models.
The approach does not rely on the asymptotic normality assumptions of previous time-interval based approaches.
Furthermore, it indicates specific areas of poor fit.
Our simulation validated the Type I error rate in a range of standard scenarios.
However, extended simulations and comparisons with other calibration approaches would provide further insight.

\

\noindent
A package to facilitate the application of the described method is available \href{https://github.com/Ben-L-Stats/BITsurv}{here.}

\newpage

\appendix

\section*{Appendix: Some practical considerations for real world applications}

\

\section{Midpoint p-values discussion} \label{midp}

When deriving test statistics and results, we will often make the assumption that midpoint p-values are uniform[0,1] under H0. 
If the validity of this approach is in question, then randomised p-values can be used, which are known to be uniform[0,1] under H0 \cite{midps}.
For this reason, the randomised p-values are superior  from this purely statistical standpoint.
However, we argue below that the midpoint p-values are more practical and that, for our applications, the uniform[0,1] assumption is often reasonable.

\

\noindent
In the case of 10 evenly-spaced intervals, as is described in example 2, the generating process is the sum of binomials. 
If the number of patients is large, one would expect these intervals to contain a reasonable number of feasible discrete outcomes, meaning we can expect the differences to be small from one discrete p-value to the next.
Because of this, when applying the midpoint p-value approach, the uniform[0,1] assumption is more appropriate.
That is, there are more possible values the midpoint p-value can take, making it behave more comparably to a uniform[0,1] distribution.

\

\noindent
In further support of midpoint p-values, we note that randomised p-values are more difficult to interpret in practice. 
Specifically, each time an analysis is rerun, the p-values are rerandomised. 
This naturally inclines an analyst to rerun the analyses to determine the level of variation in the randomised p-values and their resulting test statistic. 
This is helpful.
However, this approach leads you down the path of using the expected value of these reruns as your final quoted value. 
That is, the midpoint p-value, which is the expected value of the randomised p-value.

\section{PAVSI - practical discussion} \label{PAVSI}

In this paper, we focused primarily on the TFT test statistic as it utilises the full range of the data.
On the other hand, for PAVSI, a p-value of say 0.026 would contribute equivalently to a p-value of 0.5, which is arguably a limitation of this approach compared to the TFT.
For this reason, in general, we recommend the TFT over the PAVSI.

\ 

\noindent
Additionally, there is a caveat to using the PAVSI for the 10 evenly-spaced interval approach with midpoint p-values. 
Specifically, even if the midpoint p-values relating to each interval behave perfectly uniform[0,1] under H0, if we use a midpoint p-value to obtain the PAVSI test statistic, it can be shown analytically that the Type I error rate is 0.08614 rather than the targeted 0.05 level. 
This limitation is one of the reasons the PAVSI results are not described in the simulation section. 
Regardless, interested readers can find the PAVSI simulation results on \href{https://github.com/Ben-L-Stats/BITsurv/tree/main/Simulation/Results}{Github}, which show that when both PAVSI p-values and the interval p-values use randomised p-values, the empirical Type I error rate is close to the nominal 0.05 level.
If the reader is interested in running the scenario for randomised p-values for intervals with midpoint p-values for PAVSI, then the empirical Type I error is expected to be close to 0.08614.
Regardless of this limitation, PAVSI is included as an option in the available codes.

\ 

\noindent
In conclusion, for the censor interval approach, the PAVSI behaves well and can be well used.
For the 10 evenly-spaced interval approach with midpoint p-values, the PAVSI should be used with caution as the Type I error rate can be inflated. 
A final consideration is that in real-world analyses with the censor interval approach, extreme observations in a (likely very small) interval can render the TFT useless on account of a single extreme p-value, which results in the TFT rejecting almost every model. 
For such cases, PAVSI is a recommended alternative.
For standard cases, the TFT is generally preferred.


\end{document}